\documentclass[a4paper]{jpconf}
\usepackage{iopams}
\usepackage{graphicx}
\usepackage{amsmath,amsfonts,amssymb}
\usepackage{textpos} 

\newcommand{\be}{\begin{equation}}
\newcommand{\ee}{\end{equation}}
\newcommand{\bea}{\begin{eqnarray}}
\newcommand{\eea}{\end{eqnarray}}
\newcommand{\besub}{\begin{subequations}}
\newcommand{\eesub}{\end{subequations}}
\newcommand{\ba}{\begin{array}}
\newcommand{\ea}{\end{array}}

\newcommand{\vev}[1]{\ensuremath{\langle #1 \rangle}}
\newcommand{\GeV}{{\rm GeV}}
\newcommand{\TeV}{{\rm TeV}}

\newcommand{\Ocal}{{\cal O}}
\newcommand{\mgr}{\ensuremath{m_{3/2}}}

\newcommand{\lf}{{16 \pi^2}}

\newcommand{\tb}{{\tan \! \beta}}

\begin{document}

\begin{textblock}{3}(10.3,-1.0) \noindent DO-TH 11/04 \end{textblock}

\title{Hybrid anomaly-gravity mediation with flavor}

\author{Christian Gross}

\address{Institut f\"ur Physik, Technische Universit\"at Dortmund, 44221 Dortmund, Germany}

\ead{c.gross@physik.tu-dortmund.de}

\begin{abstract}
We study models with contributions of similar size to the soft masses from anomaly- and gravity mediation, thereby curing the tachyonic slepton problem of anomaly mediation.
A possible origin of this hybrid setup in a 5-dimensional brane world is briefly discussed.
The absence of excessive flavor violation is explained by alignment.
The gravitino can be heavy enough so that the gravitino problem of supersymmetric theories with leptogenesis is avoided.
The model has a characteristic signature:
It predicts the distinctive gaugino mass pattern of anomaly mediation and, at the same time, ${\cal{O}}(1)$ slepton mass splittings.
\end{abstract}

\section{Introduction}

While weak scale supersymmetry (SUSY) can offer solutions to some of the puzzles of the Standard Model (SM) it also introduces new challenges.
For instance, sparticle-loops can induce excessive flavor- and CP violation.
Apart from addressing this issue, a realistic supersymmetric model should also include a mechanism to generate tiny neutrino masses and the baryon asymmetry of the universe (BAU).
An elegant way to explain neutrino masses is the seesaw mechanism.
In the presence of the latter, an attractive method to explain the BAU is leptogenesis~\cite{Fukugita:1986hr}.
Since leptogenesis generically requires a reheating temperature of at least $\sim 10^9$~GeV (see e.g.~\cite{Giudice:2003jh}) the produced gravitini could destroy the successful predictions of Big Bang Nucleosynthesis (BBN) or overclose the universe~\cite{Weinberg:1982zq}.
A~straightforward way to avoid this cosmological gravitino problem is to have a gravitino mass which is large enough (i.e.~$\mgr \gtrsim 60~\TeV$, cf. e.g.~\cite{Gherghetta:1999sw}) that the gravitino decays before BBN.

This can be realized in models with anomaly mediated SUSY breaking (AMSB)~\cite{Randall:1998uk,Giudice:1998xp}, where the soft terms are loop-suppressed with respect to the gravitino mass~-- in contrast to models with Planck-scale-mediated (a.k.a. gravity-mediated) SUSY breaking (PMSB), where \mgr~generically is of the order of the soft terms.
Apart from allowing for a heavy gravitino, AMSB has the desirable feature of avoiding the SUSY flavor problem since the soft terms are not directly sensitive to UV physics.
On the other hand, AMSB has problematic aspects regarding (i) its phenomenology and (ii) its theoretical motivation:
(i)
AMSB leads to tachyonic sleptons.
There are many proposals to cure this problem (see e.g.~\cite{Randall:1998uk,Pomarol:1999ie}), none is agreed upon to be so compelling that this issue is considered settled, however.
AMSB also suffers from the $\mu$-problem.
(ii)
Being a quantum effect, AMSB generically is negligible compared to PMSB.
One thus needs to justify why PMSB is suppressed.

Here, (see~\cite{Gross:2011gj} for more details and references) we consider the possibility that the PMSB-induced soft masses are suppressed~-- in order for the gravitino to be much heavier than the \mbox{\TeV-scale~--,}  but only so much that they still are of the same size as the AMSB contribution~-- so that the tachyonic slepton problem can be avoided.
Of course, the flavor-blindness of AMSB is spoiled in this case since the PMSB contributions to the soft masses pick up flavor breaking from physics close to the Planck-scale.
In particular, the mass splittings between sleptons of different flavor and between doublet and singlet sleptons are expected to be $\Ocal (1)$.
This is not necessarily a drawback, however:
It is conceivable that the same flavor model which is needed in any case to explain the observed pattern of fermion masses and mixings also solves the SUSY flavor problem via alignment~\cite{Nir:1993mx}.
This case where physics beyond the SM is non-minimally flavor violating could actually open up possibilities to learn about the origin of flavor (see e.g.~\cite{Feng:2007ke}).

Our model has an interesting phenomenology:
The specific gaugino mass pattern of AMSB is preserved (as argued below) while we predict $\Ocal (1)$ mass splittings for the sleptons. 
Both of this can lead to characteristic signals at the LHC, providing opportunities to test the model.

\section{The model}

The gaugino masses, $A$-terms and soft masses all receive contributions from AMSB.
For the soft masses these read
\be \label{msqAMSB}
m^2|_{\rm AMSB} =  \frac{1}{2}|F_{\Phi}|^2\mu\frac{d}{d\mu}\gamma \sim |M_{\Phi}|^2 (\mp g^4-g^2 Y^2 +Y^4)\,, 
\ee
where $\gamma$ is the chiral superfield anomalous dimension, $F_{\Phi}$ is the VEV of  the conformal compensator $F$-term and $M_{\Phi}\equiv F_{\Phi}/\lf$. 
Note that $F_{\Phi} \simeq \mgr$.
The sign in~(\ref{msqAMSB}) depends on the sign of the beta-function of the corresponding gauge group and we omitted coefficients after the $\sim$~symbol.

\begin{figure}[b]
\vspace{-0pt}
  \begin{center}
    \begin{minipage}{0.36\linewidth}
      \includegraphics[width=1.0\linewidth]{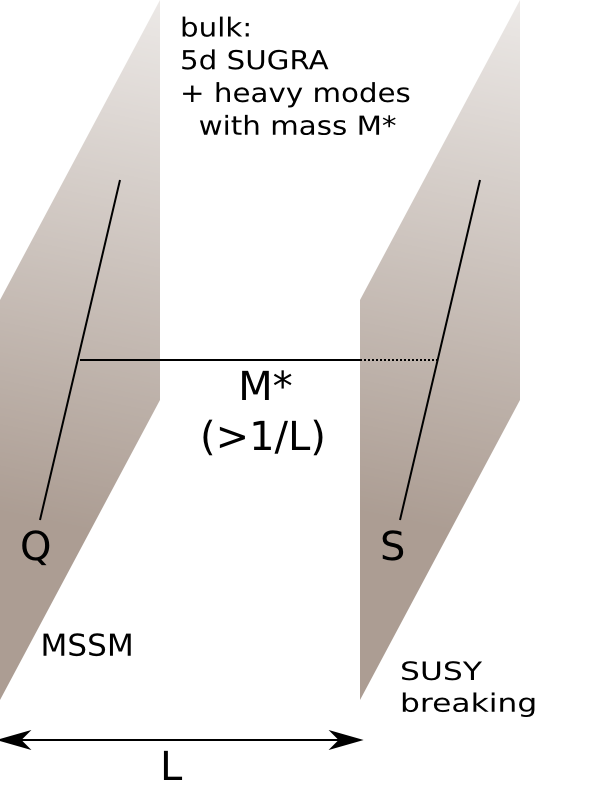}
    \end{minipage}
    \hfill
    \begin{minipage}{0.59\linewidth}
      \caption{The matter superfields, collectively denoted by $Q$, as well as the gauge- and Higgs superfields are located on the MSSM brane, which is separated from the SUSY breaking brane by a distance $L$. 
      $S$ is the hidden sector superfield with the highest $F$-term VEV, $F_S$.
      The only light fields in the bulk come from minimal 5d supergravity. 
      In addition to 5d supergravity, there may be other bulk modes, with a mass \mbox{$M_*\gtrsim1/L$}.
      These induce, by the exchange of a single propagator, effective couplings of $Q$ and $S$ which are suppressed by $e^{- M_* L}$.
      \label{fig:5d}}
    \end{minipage}
  \end{center}
  \vspace{-0pt}
\end{figure}

By contrast, the PMSB-induced contribution to the soft masses, $m^2|_{\rm PMSB}$, is of the order $F_S^2/M_*^2$ in generic models.
Here, $F_S$ is the hidden sector superfield with the highest $F$-term VEV, $F_S$, and $M_*$ is a high mass scale such as the Planck- or string scale.
Since the cancellation of the vacuum energy typically requires $F_{\Phi} \sim F_S/M_P$, the AMSB contribution is suppressed by at least a loop factor.
The desired suppression of PMSB with respect to its natural scale can be justified in 5d brane models~\cite{Randall:1998uk}, see Fig.~\ref{fig:5d}.
The 5d supergravity modes do not generate soft terms at tree level.
The effective visible-hidden sector couplings which are induced by the exchange of heavy bulk fields are exponentially suppressed and induce an operator
\be \label{operator}
\mathcal L_4 \supset \frac{e^{-M_* L}}{M_*^2} \ X_{ij} \ S \bar S Q_i \bar Q_j  \big|_{\theta^4} \,,
\ee
where $X$ is a matrix in flavor-space whose diagonal entries naturally are $\Ocal (1)$.
The term~(\ref{operator}) leads to
\be \label{PMSBsoft}
m^2|_{\rm PMSB}
\sim 
\frac{e^{-M_* L}}{M_*^2} \ |F_S|^2  \ X
\sim
|M_{\Phi}|^2\ r X \,,
\quad
\textrm{where}
\quad
r
\equiv
(\lf M_P/M_*)^2 e^{-M_* L}\,.
\ee
We used $F_S/M_P \sim F_{\Phi}$ in the last relation.
Usually it is assumed that $M_* L$ is large enough so that $r$ is tiny and $m^2|_{\rm PMSB} \ll m^2|_{\rm AMSB}$.
Here, by contrast, we assume that $r$ is of order unity, which is the case for $M_* L\simeq10+\ln(M_P^2/M_*^2)$.
One may view this as follows: The hierarchy between the mass scale of the sparticles and an $\Ocal (100\  \TeV)$~gravitino is realized by an exponential function with a suitable exponent.

Note that the PMSB contribution to the $\mu$-term, gaugino masses and $A$-terms is negligible.
This is simply because the same exponential factor suppresses a mass in these cases~-- in contrast to a mass squared in Eq.~(\ref{PMSBsoft}).
Therefore, in hybrid anomaly-gravity mediation (i) the distinctive pattern of gaugino masses of pure AMSB is maintained and, (ii), the $\mu$-problem of AMSB persists~-- we assume a viable $\mu$-term to be induced by an unspecified mechanism.

Adding the contributions from Eqs.~(\ref{msqAMSB}) and (\ref{PMSBsoft}), one obtains the chirality-preserving blocks ${\mathcal M}^{2}_{M} $ (where $M=L/R$ refers to SU(2)-doublet/singlet sleptons) of the mass matrix for the charged sleptons:
\be \label{eq:init}
{\mathcal M}^{2}_{M} 
\simeq 
|M_{\Phi}|^2 (- g_M \mathbf{1} + r  X^M ) \, ,
 \ee
where $g_{L}\equiv  (99/50) g_1^4 + (3/2) g_2^4$ and $g_{R}\equiv(198/25) g_1^4$.
Within our accuracy where the elements of $X^M$ are specified up to $\Ocal (1)$ coefficients only, the chirality-mixing blocks of the mass matrix as well as the flavor non-universal AMSB part and the $F$- and $D$-term contributions to ${\mathcal M}^{2}_{M} $ can all be neglected.
Likewise, the effects from renormalization group (RG) running between the scale $M_*$, where Eq.~(\ref{eq:init}) is defined, and the weak scale are not relevant.
(The basic reason is that sleptons do not couple to gluinos.)
We can thus employ Eq.~(\ref{eq:init}) with $g_M$ taken at the weak scale (numerically $g_{L,R} \simeq 0.3$) to deduce phenomenological consequences of our model.

\section{Aligning sleptons and explaining lepton masses/mixings}

The relative mass splitting between sleptons of different flavor and chirality is expected to be of order unity because we assume the diagonal elements of $X^M$ to have arbitrary $\Ocal (1)$ entries.
This can lead to interesting signals for slepton spectroscopy at colliders, see Sect.~\ref{sec:pred}.
It also implies that FCNCs must be suppressed by alignment only, i.e.~$\delta^M_{ij} \sim K^M_{ij}$, where $\delta^M_{ij}$ are the mass insertion parameters and $K^M_{ij}$ are the couplings of the bino and neutral wino to $M$-chiral leptons $l_i$ and sleptons $\tilde l_j$.
Since both the $g_M$ and $r$ are of order one, we have (cf.~\cite{Feng:2007ke})
\be  \label{eq:delta-eq}
\delta^{M}_{ij} \sim  \textrm{max} \left\lbrace |X^M_{ij}|, |V^M_{ij}|, |V^M_{ji}| \right \rbrace \,,
\ee
where $V^{R}, V^L$ bring the lepton Yukawa matrix $Y_E$ to diagonal form as $V^{R \dag} Y^T_E V^L$.

One way to realize sufficient alignment and explain the observed fermion masses and mixings is the Froggatt-Nielsen (FN) mechanism.
Consider as an illustration a \mbox{$U(1)_p \times U(1)_q$} symmetry under which the Higgs superfields are neutral and the lepton doublet (singlet) superfields $L_i (\bar E_i) $ have the charges
\be
L_1: (3,0),\  L_2: (1,2),\  L_3: (0,3) \, ; \qquad
\bar E_1: (3,1),\  \bar E_2: (2,-1),\  \bar E_3: (2,-3) \, .  \label{eq:FNmodel}
\ee
This yields
\be
Y_{E}
\sim
\lambda^2
\left( 
\begin{array}{ccc}
\lambda^5 & 0 & 0 \\
 \lambda^5 & \lambda^2 & 0 \\
  \lambda^5 & \lambda^2 & 1 
\end{array}
\right) 
\,,  \
X^L
\sim
\left( 
\begin{array}{ccc}
 1 & \lambda^4 & \lambda^6 \\
 \lambda^4 & 1 & \lambda^2 \\
 \lambda^6 & \lambda^2 & 1 
\end{array}
\right) , \
X^R
\sim 
\left( 
\begin{array}{ccc}
  1 & \lambda^3 & \lambda^5 \\
 \lambda^3 & 1 & \lambda^2 \\
  \lambda^5 & \lambda^2 & 1 
\end{array}
\right),
\ee
with $\lambda_p \sim \lambda_q \sim \lambda$, where $\lambda_{p,q}$ are the ratios between the spurion VEV and the heavy messenger mass scale of the FN model.
With $\lambda \sim 0.2$, this leads to realistic charged lepton masses (we assume a moderate $\tb$ here).

In this example one has \mbox{$|V^M_{ij}| \sim |V^M_{ji}| \lesssim  |X^M_{ij}|$} so that $\delta^M_{ij}\sim X^M_{ij}$.
One can now immediately check that the $\delta^M_{ij}$'s are compatible with current bounds from $l_i \to l_j \gamma$ decays (for doublet sleptons these are \mbox{$\delta^L_{12}\lesssim 6 \times10^{-4}$, $\delta^L_{13}\lesssim 0.08$, $\delta^L_{23}\lesssim 0.10$}~\cite{Ciuchini:2007ha,Gross:2011gj}), but within reach of upcoming FCNC tests~\cite{Signorelli:2003vw}.
Concerning lepton electric dipole moments (EDMs) from slepton flavor with CP violation, a rough estimate shows that a muon EDM $d_\mu$ up to $\sim 10^{-24} \, {\rm e\, cm}$ is possible, which is well below the current bound $d_{\mu}=(-0.1\pm0.9) \times 10^{-19} \, {\rm e\, cm}$~\cite{Bennett:2008dy}, but within reach of the proposed activities to measure $d_\mu$ as low as $5 \times 10^{-25} \, {\rm e\, cm}$~\cite{Adelmann:2010zz}. 
The electron EDM $d_e$ has a much stronger experimental bound but at the same time the relevant mass insertion parameters are suppressed even further so that $d_e$ is also below the current bound.

We assume that the neutrino masses are generated by the seesaw mechanism (with neutral Majorana neutrinos $N_i$).
The charges~(\ref{eq:FNmodel}) are chosen such that the neutrino sector is anarchical,
\bea \label{neutrinoanarchy}
(Y_{N})_{ij}&\sim&
\lambda^{n_{\nu}} , 
\qquad
(m_\nu)_{ij} \sim
 \lambda^{2 n_{\nu}}
\vev{H^{0}_{u}}^2/\hat M_{R} \,,
 \quad \forall\, i,j \,,
\eea
with $n_{\nu}=3$.
One may worry that the Yukawa matrix $Y_N$ induces excessive flavor violation via RG evolution~\cite{Borzumati:1986qx}.
One can show that this effect is $\delta X^L_{ij}\sim 0.1\,  \lambda^{2 n_{\nu}}$ which is negligible within our accuracy.
Note also that, in order to arrive at a neutrino mass scale of $\sim 0.1$~eV, the $N_i$ should have masses around $\hat M_{R}\sim 10^{10}\ \GeV$, which is compatible with leptogenesis (cf. e.g.~\cite{Davidson:2002qv,Giudice:2003jh}).

Within the framework of FN symmetries, there exists a lower limit on the magnitude of the $\delta^L_{ij}$'s, for the following reason.
The charges of the $L_i$ determine both $X^L$ and $Y_N$. 
The latter should not be suppressed too much, otherwise the seesaw scale would be too low.
Therefore, the possible suppression of the $X^L_{ij}$'s and hence $\delta^L_{ij}$'s is limited.
Nevertheless, it is possible within \mbox{$U(1)_p \times U(1)_q$} models to construct examples (see~\cite{Gross:2011gj}) which produce a precise enough alignment that no signal would be seen even in planned future rare decay measurements.

\section{Signatures of the model} \label{sec:pred}

\begin{figure}[b]
\vspace{-10pt}
  \begin{center}
    \begin{minipage}{0.50\linewidth}
      \includegraphics[width=1.0\linewidth]{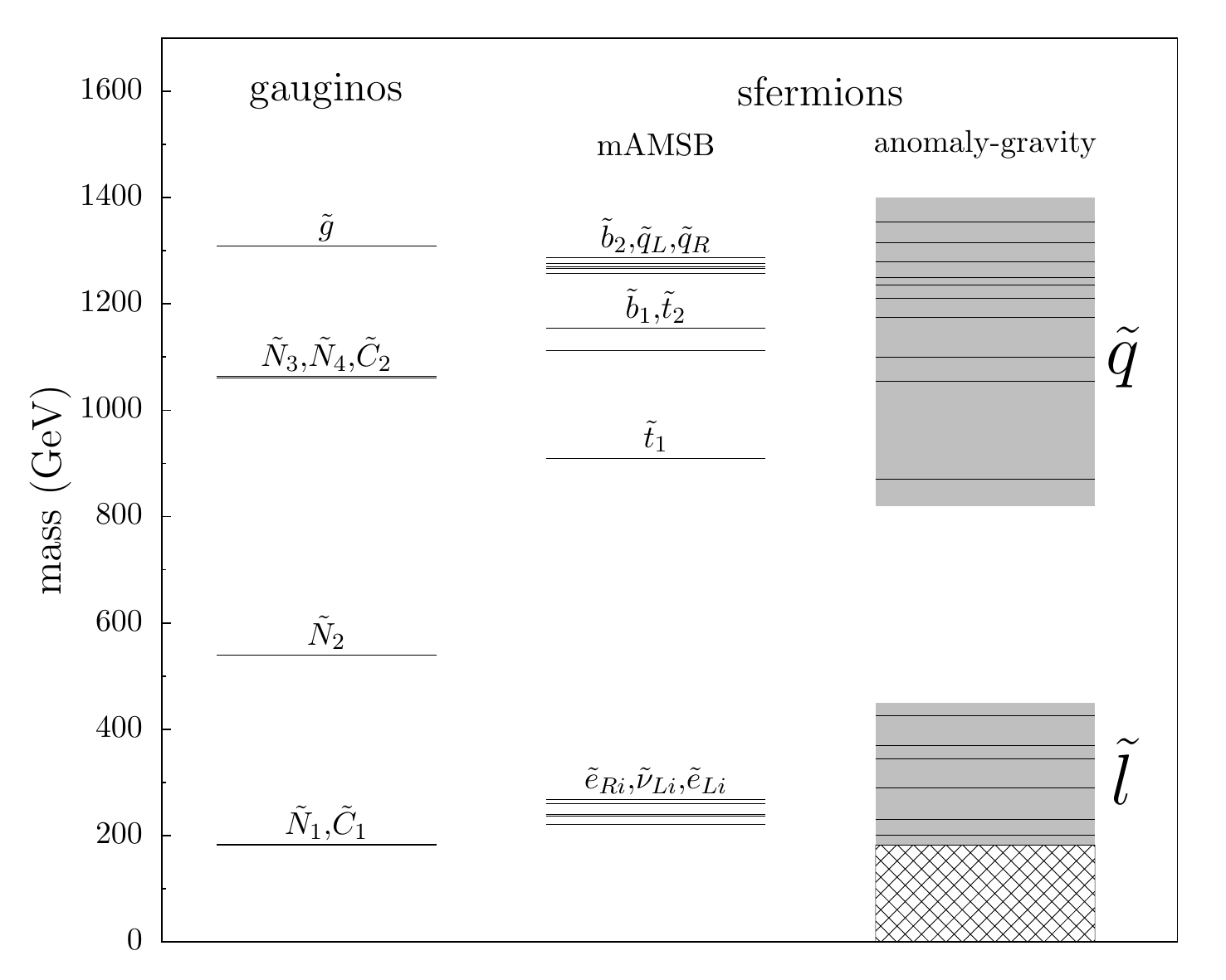}
    \end{minipage}
    \hfill
    \begin{minipage}{0.45\linewidth}
      \caption{Schematic plot (taken from~\cite{Gross:2011gj}) of a sample sfermion spectrum in hybrid AMSB-PMSB (3rd column), compared to an mAMSB spectrum (2nd column) with universal scalar mass uplift $m_0= 350 \ \GeV$ and $\tb =5$, $m_{3/2}=60 \ \TeV$, $\mu >0$.
The gaugino masses (1st column) are as in pure AMSB (and mAMSB).
The crosshatched band is disfavored phenomenologically.
\label{fig:spectrum}}
    \end{minipage}
  \end{center}
    \vspace{-12pt}
\end{figure}

The model has a characteristic signature:
The unique pattern of gaugino masses of AMSB (see Fig.~\ref{fig:spectrum}, 1st column) is preserved, since, as discussed above, the PMSB contribution to the gauginos is negligible.
The gaugino masses thus have a ratio $|M_1| : |M_2| : |M_3|$ of $3:1:7$, yielding in particular an almost degenerate wino-like lightest neutralino $\tilde N_1$ and chargino $\tilde C_1$. 
A distinctive signal is the soft pion in the decay $\tilde C_1^{\pm} \to \tilde N_1 \pi^{\pm}$~\cite{Feng:1999fu,Gherghetta:1999sw}.

At the same time, the model predicts $\Ocal(1)$ slepton mass splittings~-- both between sleptons of different flavor and between singlet and doublet sleptons (as illustrated in Fig.~\ref{fig:spectrum}, 3rd column).
This could possibly be probed at the LHC by same-flavor dilepton edge measurements with missing energy in $\tilde N_2 \to l \tilde l^* , \bar  l \tilde l \to \bar l l \tilde N_1$ cascades.
The position of the edge depends on the mass of the intermediate slepton.
A rough estimate shows that the rates for singlets and doublets are of a similar order of magnitude, related to the bino nature of $\tilde N_2$.
It thus seems not unlikely that the edges would be distinguishable at the LHC.
The selectron-smuon mass difference could be obtained by comparing the respective kinematical edges (cf. e.g. the mSUGRA study~\cite{Allanach:2008ib}).

The squark mass squared splittings are expected to be $\Ocal(10\%)$ only, due to the large flavor-universal RG-effect of the gluinos.
This leads to $D$-$\bar D$ mixing close to the experimental limit~\cite{Feng:2007ke} and a lower bound on hadronic EDMs~\cite{Altmannshofer:2010ad}.

As far as low-energy lepton flavor experiments are concerned, it is well possible that a signal is around the corner, such as is the case in the example we presented above for $\mu \to e \gamma$.
A non-observation of any signal even in future FCNC experiments would however not falsify the framework of hybrid anomaly-gravity mediation.

\section{Conclusions}

Models where SUSY is broken by a combination of anomaly- and gravity mediation can be motivated in 5d brane worlds.
The flavorful gravity mediated contribution to the slepton masses avoids the tachyonic slepton problem of AMSB.
From a model-building perspective, a virtue is the possibility to have leptogenesis without the cosmological gravitino problem.
From a phenomenological perspective, the model is interesting because it has a rather special signature: AMSB-like gaugino masses and~-- since the SUSY flavor problem is solved by alignment only~-- order one slepton mass splittings.

It would be worthwhile to elaborate the model further by, for instance, including a mechanism to generate a viable $\mu$-term and a mechanism to stabilize the branes at a distance that leads to $r \sim 1$, cf. Eq.~(\ref{PMSBsoft}).
Also, the phenomenological implications of the model deserve further study.
In particular, it would be interesting to explore in more detail the prospects of measuring the large slepton mass splittings at the LHC.

\ack
I would like to thank Gudrun Hiller for collaboration on the work presented here.
This work is supported in part by the {\it Bundesministerium f\"ur Bildung und Forschung (BMBF)}.



\section*{References}
\begin{thebibliography}{99}

\bibitem{Fukugita:1986hr}
Fukugita~M and Yanagida~T 1986
  {\it Phys.\ Lett.\ }B {\bf 174} 45
  
\bibitem{Giudice:2003jh}
  Giudice~G~F, Notari~A, Raidal~M, Riotto~A and Strumia~A 2004
{\it  Nucl.\ Phys.}\  B {\bf 685} 89

\bibitem{Weinberg:1982zq}
  Weinberg~S 1982
  {\it Phys.\ Rev.\ Lett.}\  {\bf 48} 1303
  
\bibitem{Gherghetta:1999sw}
  Gherghetta~T, Giudice~G~F and Wells~J~D 1999
  {\it Nucl.\ Phys.}\ B {\bf 559} 27
  
\bibitem{Randall:1998uk}
 Randall~L and Sundrum~R 1999
  {\em Nucl. Phys.}\  B\ {\bf 557} 79

\bibitem{Giudice:1998xp}
  Giudice~G~F, Luty~M~A, Murayama~H and Rattazzi~R 1998
  {\em JHEP} {\bf 9812} 027
    
\bibitem{Pomarol:1999ie}
  Pomarol~A and Rattazzi~R 1999
  {\it JHEP} {\bf 9905} 013
  
\bibitem{Gross:2011gj}
  Gross~C and Hiller~G 2011
  {\it Preprint} arXiv:1101.5352

\bibitem{Nir:1993mx}
Nir~Y and Seiberg~N 1993
{\it Phys.\ Lett.\ }B {\bf 309} 337

\bibitem{Feng:2007ke}
Feng~J~L, Lester~C~G, Nir~Y and Shadmi~Y 2008
{\it Phys.\ Rev.\ }D {\bf 77} 076002

\bibitem{Ciuchini:2007ha}
  Ciuchini~M, Masiero~A, Paradisi~P, Silvestrini~L, Vempati~S~K and Vives~O 2007
 {\it  Nucl.\ Phys.\ }B {\bf 783} 112

\bibitem{Signorelli:2003vw}
  Signorelli~G 2003
{\it   J.\ Phys.\ }G {\bf 29} 2027
  
\bibitem{Bennett:2008dy}
  Bennett~G~W {\it et al.} [Muon (g-2) Collaboration] 2009
{\it   Phys.\ Rev.\ }D {\bf 80} 052008
  
\bibitem{Adelmann:2010zz}
  Adelmann~A, Kirch~K, Onderwater~C~J~G and Schietinger~T 2010
  {\it J.\ Phys.\ }G {\bf 37} 085001
  
\bibitem{Borzumati:1986qx}
  Borzumati~F and Masiero~A 1986
{\it   Phys.\ Rev.\ Lett.}\  {\bf 57} 961

\bibitem{Davidson:2002qv}
  Davidson~S and Ibarra~A 2002
{\it   Phys.\ Lett.}\  B {\bf 535} 25
  
\bibitem{Feng:1999fu}
  Feng~J~L, Moroi~T, Randall~L, Strassler~M and Su~S 1999
{\it   Phys.\ Rev.\ Lett.}\  {\bf 83} 1731

\bibitem{Allanach:2008ib}
Allanach~B~C, Conlon~J~P and Lester~C~G 2008
{\it  Phys.\ Rev.}\ D  {\bf 77} 076006
    
\bibitem{Altmannshofer:2010ad}
  Altmannshofer~W, Buras~A~J and Paradisi~P 2010
{\it   Phys.\ Lett.\ }B {\bf 688} 202

\end{thebibliography}
\end{document}